\documentclass[aps,prl,reprint,letterpaper,amsmath,amssymb,superscriptaddress,twocolumn,preprintnumbers,longbibliography]{revtex4-1}

\usepackage{graphicx}
\usepackage{times}
\usepackage{bm}
\usepackage[dvipsnames]{xcolor}
\usepackage{ulem}
\usepackage{amsmath}
\usepackage{float}

\def\emph{\textit}

\def\be{\begin{equation}}
\def\ee{\end{equation}}
\def\bea{\begin{eqnarray}}
\def\eea{\end{eqnarray}}

\newcount\mycounter

\begin{document}


\title{Coherent spin dynamics of electrons and holes in CsPbBr$_3$ perovskite crystals}

\author{Vasilii~V. Belykh}
\email{vasilii.belykh@tu-dortmund.de}
\affiliation{Experimentelle Physik 2, Technische Universit\"{a}t Dortmund, D-44221 Dortmund, Germany}
\author{Dmitri~R. Yakovlev}
\email{dmitri.yakovlev@tu-dortmund.de}
\affiliation{Experimentelle Physik 2, Technische Universit\"{a}t Dortmund, D-44221 Dortmund, Germany}
\affiliation{Ioffe Institute, Russian Academy of Sciences, 194021 St. Petersburg, Russia}
\author{Mikhail~M. Glazov}
\affiliation{Ioffe Institute, Russian Academy of Sciences, 194021 St. Petersburg, Russia}
\author{Philipp~S. Grigoryev}
\affiliation{Spin Optics Laboratory, St. Petersburg State University, 199034 St. Petersburg, Russia}
\author{Mujtaba~Hussain}
\affiliation{Centre for Micro and Nano Devices, Department of Physics,
COMSATS University, 44000 Islamabad, Pakistan}
\author{Janina~Rautert}
\affiliation{Experimentelle Physik 2, Technische Universit\"{a}t Dortmund, D-44221 Dortmund, Germany}
\author{Dmitry~N.~Dirin}
\affiliation{Laboratory of Inorganic Chemistry, Department of Chemistry and Applied Biosciences,  ETH Z\"{u}rich, CH-8093 Z\"{u}rich, Switzerland}
\author{Maksym~V.~Kovalenko}
\affiliation{Laboratory of Inorganic Chemistry, Department of Chemistry and Applied Biosciences,  ETH Z\"{u}rich, CH-8093 Z\"{u}rich, Switzerland}
\affiliation{Laboratory for Thin Films and Photovoltaics, Empa-Swiss Federal Laboratories for Materials Science and Technology, CH-8600 D\"{u}bendorf, Switzerland}
\author{Manfred~Bayer}
\affiliation{Experimentelle Physik 2, Technische Universit\"{a}t Dortmund, D-44221 Dortmund, Germany}
\affiliation{Ioffe Institute, Russian Academy of Sciences, 194021 St. Petersburg, Russia}

\date{October 9, 2018}

\maketitle

\textbf{The lead halide perovskites demonstrate huge potential for optoelectronic applications, high energy radiation detectors, light emitting devices and solar energy harvesting~\cite{Stoumpos2013, Manser2016, Li2016, Dursun2016}.  Those materials exhibit strong spin-orbit coupling enabling efficient optical orientation of carrier spins~\cite{Giovanni2015,Odenthal2017} in perovskite-based devices with performance controlled by a magnetic field~\cite{Zhang2015}. Perovskites are promising for spintronics~\cite{Sun2016,Yu2016} due to  substantial bulk and structure inversion asymmetry~\cite{Kepenekian2015,Kim2014,Niesner2016}, however, their spin properties are not studied in detail. Here we show that elaborated time-resolved spectroscopy involving strong magnetic fields can be successfully used for perovskites. We perform a comprehensive study of high-quality CsPbBr$_3$ crystals by measuring the exciton and charge carrier $g$-factors, spin relaxation times and hyperfine interaction of carrier and nuclear spins by means of coherent spin dynamics. Owing to their ``inverted'' band structure, perovskites represent appealing model systems for semiconductor spintronics exploiting the valence band hole spins, while in conventional semiconductors  the conduction band electrons are considered for spin functionality.}

Semiconductor spintronics is an intense research field covering the whole variety of spin-dependent phenomena and numerous experimental techniques, which allow one to study the spin structure and spin dynamics in different materials and their nanostructures. Optical techniques with time- and polarization resolution and application of magnetic field are widely used for that. Despite the great recent interest to various perovskite materials, including two-dimensional perovskites and colloidal nanocrystals, spin studies are at the very beginning here. It has been demonstrated, however, that experimental approaches like optical orientation~\cite{Nestoklon2018}, spin polarization induced by magnetic field~\cite{Zhang2015,Canneson2017}, pump-probe Faraday rotation~\cite{Giovanni2015,Odenthal2017}, and single dot spectroscopy in magnetic field~\cite{Fu2017,Isarov2017,Ramade2018} are working well for perovskites and their nanostructures. The fine structure of neutral and charged excitons has been addressed, including their spin dynamics. Recently, it has been shown that the combination of spin-orbit and exchange interaction in perovskite nanocrystals may result in an unusual ordering of the exciton fine structure levels with an optically active triplet ground state~\cite{Becker2018}.

%

Here we report spin-dependent phenomena in CsPbBr$_3$ perovskite crystals of high structural and optical quality, as confirmed by sharp exciton resonances in reflectivity and emission spectra. We focus on the coherent spin dynamics in external magnetic fields at cryogenic temperatures studied by optical techniques based on the pump-probe time-resolved Kerr rotation. We measure the transverse and longitudinal spin relaxation times of electrons and holes and their dependencies on magnetic field and temperature. We evaluate the exciton, electron and hole $g$-factors including their signs and spread. Polarizing the nuclear spins dynamically via optically oriented carriers, we address hyperfine interaction effects and find the dominant role of the holes in them, which is in agreement with our model considerations.
\begin{figure*}
    \centering
\includegraphics[width=1.5\columnwidth]{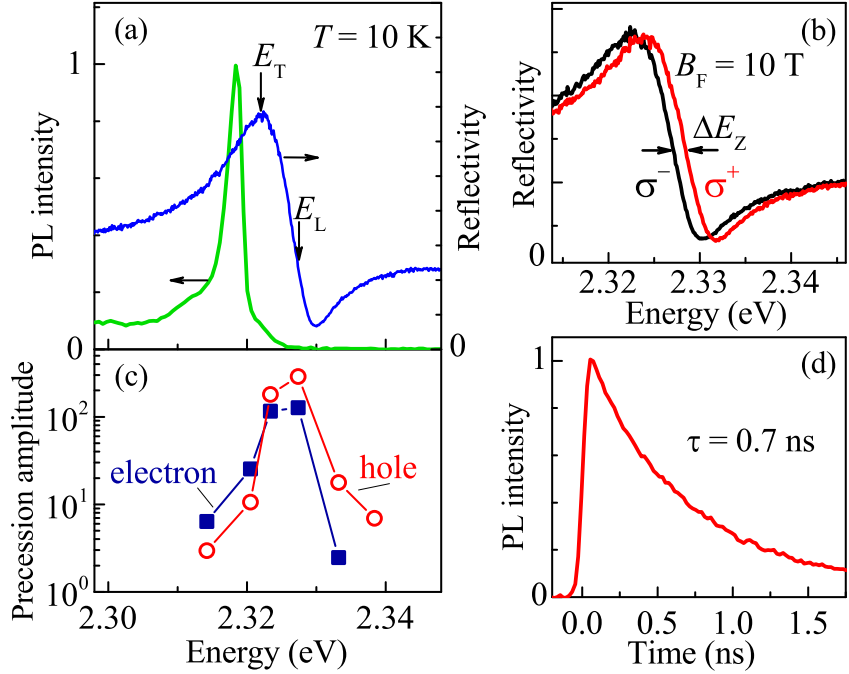}
\caption{\textbf{Photoluminescence and reflectivity of CsPbBr$_3$ perovskite crystal.} \textbf{a}, Photoluminescence (green line, excitation energy at 2.376~eV) and reflectivity (blue line) spectra. Energies for longitudinal ($E_\text{L}$) and transverse ($E_\text{T}$) exciton-polaritons are marked by arrows.
\textbf{b}, Reflectivity spectra measured for opposite circular polarizations in longitudinal magnetic field $B_\text{F} = 10$~T.
\textbf{c}, Spectral dependence of spin precession amplitude of electrons (solid squares) and holes (open circles) in transverse magnetic field $B_\text{V} = 0.5$~T.
\textbf{d}, Exciton recombination dynamics measured at 2.318~eV excitation photon energy with streak-camera. $T=10$~K.
}
    \label{fig:Spectra}
\end{figure*}

We start with optical characterization of the CsPbBr$_3$ perovskite crystal at a low temperature of 10~K. The reflectivity spectrum shown in Fig.~\ref{fig:Spectra}a by the blue line demonstrates a strong exciton-polariton resonance with transverse and longitudinal energies of $E_\text{T} = 2.3220$~eV and $E_\text{L}= 2.3274$~eV (Methods). In a magnetic field of $B_\text{F} = 10$~T the reflectivity spectra in the two opposite circular polarizations show an exciton Zeeman splitting of $\Delta E_\text{Z} = 1.32$~meV (Fig.~\ref{fig:Spectra}b), which corresponds to the exciton $g$-factor $g_\text{X} = \Delta E_\text{Z}/(\mu_\text{B} B_{\rm F}) = 2.35$, where $\mu_\text{B}$ is the Bohr magneton.
In the photoluminescence (PL) spectrum (the green line in Fig.~\ref{fig:Spectra}a) the narrow exciton peak at 2.318~eV has a small Stokes shift of 4~meV from $E_\text{T}$. The exciton has a lifetime of 0.7~ns measured by time-resolved PL (Fig.~\ref{fig:Spectra}d).  The PL band at the lower energy side of the exciton presumably arises due to bound excitons.

\begin{figure*}
    \centering
\includegraphics[width=1.5\columnwidth]{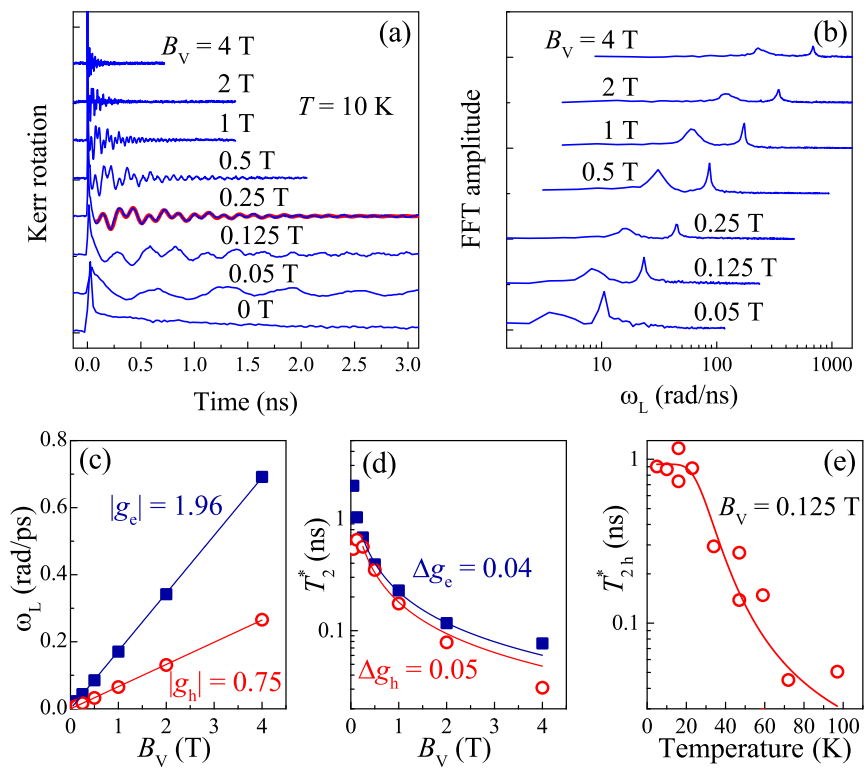}
\caption{\textbf{Coherent spin dynamics in transverse magnetic field.}
\textbf{a}, Kerr rotation dynamics in CsPbBr$_3$ crystal at different magnetic fields. Magenta thick line for  $B_\text{V} = 0.25$~T is fit to the experimental data with two decaying oscillatory functions (Methods).
\textbf{b}, Fast Fourier transform spectra of spin dynamics traces from panel \textbf{a}.
\textbf{c}, Magnetic field dependencies of electron (squares) and hole (circles) Larmor frequencies. Lines show linear fits to data.
\textbf{d}, Magnetic field dependencies of electron (squares) and hole (circles) spin dephasing times. Lines show reciprocal to $B_\text{V}$ fits to data with evaluated spread of $g$-factors. \textbf{a-d}, $T=10$~K.
\textbf{e}, Temperature dependence of hole spin dephasing time. Line is fit with activation dependence having energy parameter $\Delta E \approx 14$~meV.
}
    \label{fig:Kin}
\end{figure*}

The coherent spin dynamics of carriers is measured by the time-resolved pump-probe Kerr rotation. Figure~\ref{fig:Kin}a shows the spin dynamics at different magnetic fields $B_\text{V}$ applied perpendicular to the pump and probe beams (Voigt geometry). The oscillating signals result from the Larmor spin precession of the carrier spin polarization about the magnetic field with frequency $\omega_{\text{L},e(h)} = |g_{e(h)}| \mu_\text{B} B_\text{V} / \hbar$.~\cite{Glazov2012,Yakovlev_Ch6} Here $g_{e}$ and $g_{h}$ are the electron and hole $g$-factors, respectively. The signal precession is seemingly aperiodic, which is due to the presence of two frequencies, as evidenced from two peaks in the fast Fourier transform (FFT) spectra of the spin dynamics (Fig.~\ref{fig:Kin}b). The two frequencies increase linearly with magnetic field (Fig.~\ref{fig:Kin}c) and correspond to $g$-factors of $|g_\text{e}| = 1.96$ and $|g_\text{h}| = 0.75$. This assignment of the $g$-factors to electrons and holes is based on the following arguments. First, the exciton contribution is excluded as it should be characterized by a $g$-factor of $g_\text{X} = 2.35$ extracted directly from the Zeeman splitting in reflectivity. Second, both $\omega_\text{L}(B_\text{V})$ dependencies can be extrapolated to zero frequency for vanishing field, i.e. no contribution of an exciton exchange splitting is seen~\cite{Ramade2018}. Also, the spin dephasing times for both frequencies are longer than the 0.7~ns exciton lifetime. In the perovskites $g_\mathrm{X} = g_e + g_h$ (Section S4 in Supplementary information) and we found experimentally that $g_\text{X} > 0$ and $|g_{e}|, |g_{h}| < g_\mathrm{X}$. Therefore, we conclude that in the studied material $g_e>0$ and $g_h>0$. A specifics of the perovskite band structure, compared to common II-VI and III-V semiconductors, is the strong renormalization of the hole $g$-factor compared to the electron one~\cite{Yu2016}. This allows us to assign the 1.96 $g$-factor to the electron, and the smaller $g$-factor to the hole.

The electron and hole spin signals are maximal for laser energies close to the exciton-polariton resonance (Fig.~\ref{fig:Spectra}c) due to efficient spin initialization and detection in resonance. The dephasing time of the spin precession, $T_2^*$, shortens with increasing $B_\text{V}$ (Fig.~\ref{fig:Kin}d), which is related to the spread of $g$-factor values, $\Delta g$. It can be described by $1/T_2^* = 1/\tau_\text{s}+\Delta g \mu_\text{B} B_\text{V} / \hbar$, where $\tau_\text{s}$ is the spin lifetime at zero field. A fit with this equation gives $\Delta g_\text{e} = 0.04$, $\tau_{\mathrm{s},e} =2$~ns, $\Delta g_\text{h} = 0.05$,  and $\tau_{\mathrm{s},h} =1$~ns.

We suggest that the spin dynamics is contributed by resident electrons and holes localized in CsPbBr$_3$ at spatially separated locations. The resident carriers can be provided by unintentional doping in solution-grown crystals~\cite{Dirin2016} or by photogeneration, which is in line with the remarkable photovoltaic properties of perovskites. The lifetime of these carriers is significantly longer than the laser repetition period of 13.1~ns~\cite{Stoumpos2013,He2018}. Therefore, the spin polarization can be accumulated from many subsequent pump pulses\cite{Yakovlev_Ch6}. The mechanism of spin coherence generation for resident carriers is the same as in semiconductor quantum wells and quantum dots, see Section S2 in Supplementary information and Refs.~\onlinecite{Yakovlev_Ch6,Glazov2012}. As we show below, the established experimental approaches for spin dynamics studies in conventional semiconductor nanostructures are also suitable for the perovskites.

\begin{figure*}
    \centering
\includegraphics[width=1.5\columnwidth]{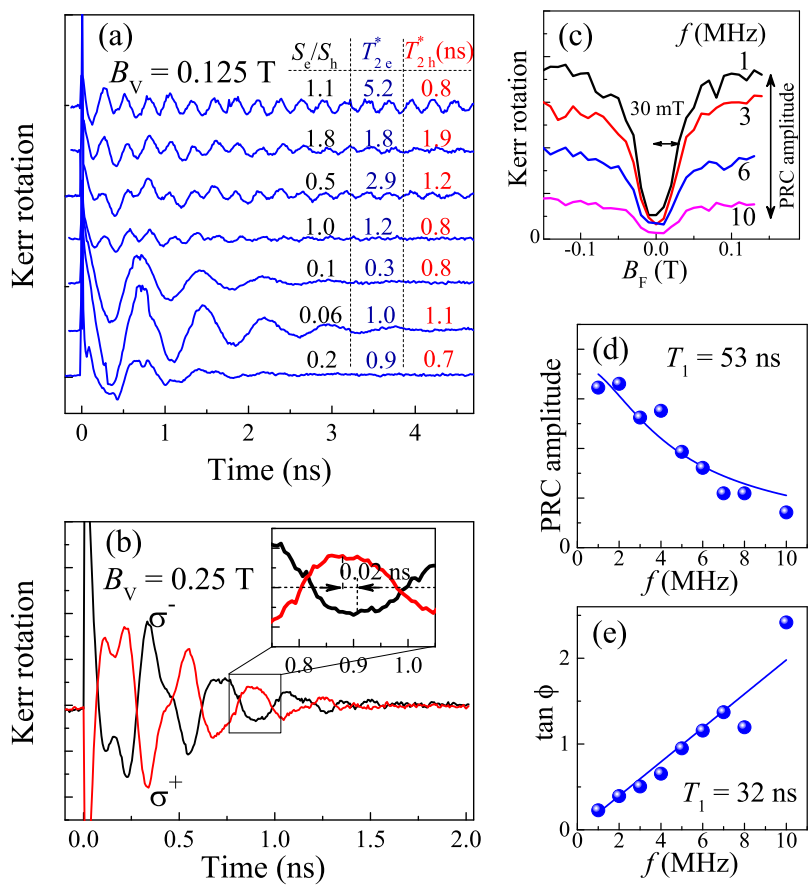}
\caption{\textbf{Carrier-nuclei hyperfine interaction and evaluation of longitudinal spin relaxation time $T_1$.}
\textbf{a}, Spin dynamics at different positions on sample, $B_\text{V} = 0.125$~T and $T=10$~K.
\textbf{b}, Dynamics of Kerr rotation for different circular polarizations of pump pulses. Inset illustrates phase shift acquired for hole spin precession. $T = 5$~K. Pump is tilted from normal incidence by an angle of 15$^{\circ}$.
\textbf{c}, Polarization recovery curves (PRCs): dependencies of Kerr rotation signal on longitudinal magnetic field at time delay $\Delta t = 13$~ns, measured for different pump modulation frequencies. $T = 2$~K.
\textbf{d}, Modulation frequency dependence of PRC amplitude. Line is fit to data with Eq.~\eqref{eq:amp} giving $T_1 = 53$~ns.
\textbf{e}, Modulation frequency dependence of $\tan\phi$, where $\phi$ is phase retardation of Kerr rotation signal with respect to pump modulation. Line is linear fit $\tan\phi = 2\pi f T_1$ with $T_1 = 32$~ns.
  }
    \label{fig:PRC}
\end{figure*}

In the studied CsPbBr$_3$ crystal the relative amplitudes of the electron and hole signals vary with laser spot position (Fig.~\ref{fig:PRC}a). This evidences the inhomogeneous spatial distribution of the resident carriers and confirms their localization. Note that the hole spin dephasing time at weak magnetic fields ($T_{2,h}^* \approx \tau_{\mathrm{s},h} \approx 1$~ns) is not sensitive to the spot position.
By contrast, the electron dephasing time $T_{2,e}^*$ has a stronger dependence on the spot position reaching up to 5.2~ns, which exceeds by almost an order of magnitude the exciton lifetime of 0.7~ns.

The electron and hole spin dephasing times are almost constant at low temperatures $T \le 15$~K. At higher temperatures the electron spin precession is not detectable due to the abrupt shortening of $T_{2,e}^*$. Hole spin dynamics can be measured up to 100~K and $T_{2,h}^*$ smoothly decreases with increasing temperature (Fig.~\ref{fig:Kin}e). This behavior can be described by an activation-type dependence: $1/T_{2}^*(T) = 1/T_{2}^*(0) + w \exp(-\Delta E / k_\text{B} T)$. Here $T_2^*(0)$ is the spin dephasing time at zero temperature, $w$ is a phenomenological pre-factor, and $k_\text{B}$ is the Boltzmann constant. The activation energy $\Delta E \approx 14$~meV and $w \approx 160$~ns$^{-1}$ are evaluated from the fit. This behavior can be related to either hole delocalization or to a spin-flip process mediated by LO phonons, whose energy in CsPbBr$_3$ is $\sim 18$~meV.~\cite{Pashuk1981}

In semiconductors the spin dynamics of localized carriers is controlled by the carrier hyperfine interaction with the nuclear spins~\cite{OO1984}, as the quenching of the orbital motion suppresses the spin-orbit coupling effects. The nuclear spins experience the Knight field from the spin-polarized carriers and, in turn, the carriers experience the Overhauser field induced by the nuclei. To examine nuclear effects, we intentionally polarize the nuclear spins by setting the circular polarization of the pump beam and  tilting it by an angle of 15$^\circ$ from the normal to the sample surface. This provides a non-zero projection of the charge carrier spin polarization onto the magnetic field. Flip-flop hyperfine processes transfer the carrier spin polarization to the nuclear spin system. This gains a dynamical nuclear polarization $\mathbf I$ (Section S3 in Supplementary information) and induces an Overhauser field $\mathbf B_{{\rm N},e(h)} = A_{e(h)} \mathbf I/(g_{e(h)}\mu_B)$, that adds up to the external field. It  changes the frequency of the carrier spin precession. The direction of the Overhauser field is determined by the pump helicity.

Figure~\ref{fig:PRC}b shows that the spin beats for opposite pump polarizations acquire a small but detectable relative phase shift, which increases with delay time, evidencing a difference in the spin precession frequencies. An accurate fit of the experimental data shows that the Overhauser field acting on the holes $|B_{\mathrm N,h}| =3.1 \pm 0.5$~mT is three times larger than that on the electrons $|B_{\mathrm N,e}| = 1.0 \pm 0.8$~mT. This result may seem surprising compared with the widely studied III-V and II-VI semiconductors where the hyperfine coupling is dominated by the conduction band electrons~\cite{Gryncharova1977,Chekhovich2013}. Our theoretical analysis demonstrates that the stronger hyperfine coupling for the valence band holes compared to the conduction band electrons is a particular feature of perovskites such as CsPbBr$_3$ (Section S3 in Supplementary information). We estimate that the hole hyperfine coupling by the Fermi contact interaction with the $^{207}$Pb isotopes with $I=1/2$ amounts to $A_{h}\approx 20$~$\mu$eV. The somewhat weaker dipole-dipole interaction of the conduction band electron with the $^{79}$Br and $^{81}$Br isotopes with $I=3/2$ gives $A_{e} \approx 7$~$\mu$eV. These estimates demonstrate that the dynamical nuclear polarization is far from $100$~\%, most probably due to spin relaxation processes unrelated to the hyperfine coupling, see Supplementary information for details.

At low temperatures and weak magnetic fields the spin dephasing time $T_2^*$ of localized charge carriers is mainly contributed by the static fluctuations of the nuclear Overhauser field. Application of a longitudinal magnetic field $B_\text{F}$ (Faraday geometry) parallel to the initial spin polarization suppresses the transverse fluctuations and stabilizes carrier spins against the influence of random nuclear fields. Insight into the effect can be obtained  from polarization recovery curves (PRC)~\cite{Heisterkamp2015} (Fig.~\ref{fig:PRC}c). The polarization recovery for alternating optical orientation allows us also to evaluate the longitudinal spin relaxation time $T_1$ of the charge carriers via the spin inertia method~\cite{Heisterkamp2015} (Methods).  Typical polarization recovery curves (PRC) show an increase of the spin polarization with increasing magnetic field exhibiting a half width at half maximum of 30~mT and saturation with growing $B_\text{F}$ (Fig.~\ref{fig:PRC}c). The saturation level decreases with increasing the pump modulation frequency $f$ from 1 to 10~MHz. Fitting of the dependence of the PRC amplitude on $f$ with Eq.~\eqref{eq:amp} (Methods) gives $T_1 = 53 \pm 9 $~ns (Fig.~\ref{fig:PRC}d). The measurement of the frequency dependence of the phase retardation, $\phi$, of the spin polarization signal with respect to that of the pump modulation allows us to evaluate $T_1 = 32 \pm 2$~ns with $\tan\phi = 2\pi f T_1$ (Fig.~\ref{fig:PRC}e). The different $T_1$ values evidence a non-monoexponential decay of the spin polarization (Methods). We also find that $T_1$ is constant for a temperature increase from 2 to 10~K, and then strongly decreases to a few ns for $T \gtrsim 20$~K, presumably due to the same activation process as relevant for $T_{2,h}^*$ (Fig.~\ref{fig:Kin}e).
The analysis of the PRC and spin inertia signals provides an estimate of the nuclear field fluctuations $\delta B_\text{N} \approx 6.6$~mT and the hole correlation time at the localization site $\tau_c\approx 2.1$~ns (Section S5 in Supplementary information).

In summary, we have demonstrated that spin phenomena show up as prominent features in the optical properties of perovskites, even though they have remained largely unexplored so far. Fortunately, the methodology established for other semiconductors can be transferred to the perovskites. In particular we have elaborated the importance of the nuclear spins in these phenomena, which may be used as additional resource, for example, for establishing a long-lived spin memory. Based on our data, one may seek also for coherent spin control in perovskites.

\subsection*{Acknowledgements}

V.V.B., D.R.Y., P.S.G., J.R., and M.B. acknowledge  support  of  the  Deutsche Forschungsgemeinschaft through the International Collaborative Research Centre TRR 160 (projects A1 and B2). The stay of M.H. in Dortmund was supported by the DAAD (project EXCIPLAS, headed by Arshad Bhatti on the Pakistani side). P.S.G. acknowledges the Saint-Petersburg State University for a grant 11.34.2.2012. M.M.G. was partially supported by RFBR and RF President Grant No. MD-1555.2017.2. M.M.G. and M.B. were also supported by RF Government Grant No. 14.Z50.31.0021. We thank M.A. Semina for providing results of exciton-polariton modeling.





\subsection*{Methods}

\textbf{Growth of CsPbBr$_3$ samples.}
Single crystals of CsPbBr$_3$ were grown as reported elsewhere with slight modifications~\cite{Dirin2016}. First, CsBr and PbBr$_2$ were dissolved in dimethyl sulfoxide at concentrations of 0.5~M and 1~M, respectively and the resulting solution (2~mL) was filtered through PTFE filter (0.2~$\mu$m). 2~mL of the cyclohexanol solution in N,N-dimethylformamide (5.1~g in 9.1~g, respectively) were added and the resulting mixture was heated in an oil bath to 70$^\circ$C and then slowly ($\sim 0.05-0.1$~$^\circ$C/min) to 105$^\circ$C. After $\sim 12$ hours of growth, the obtained crystals were taken out of the solution and quickly loaded into a vessel with hot (100$^\circ$C) N,N-dimethylformamide. This vessel was slowly ($\sim 25$~$^\circ$C/hour) cooled down to $\sim 50^\circ$C. After that the crystals were isolated, wiped with filter paper and dried. The obtained rectangular CsPbBr$_3$ is crystallized in the orthorhombic modification. The crystals have a one selected (long) direction along the c-axis and two nearly identical directions along the [110] and [110] axes.

\textbf{Reflectivity and photoluminescence characterization.}
For optical experiments the CsPbBr$_3$ sample was placed in a liquid-helium-cooled cryostat, where the sample temperature, $T$, was varied from 2 up to 100~K. The cryostat was equipped with a superconducting split-coil solenoid generating external magnetic fields up to 10~T, which were applied either parallel to the light wave vector in the Faraday geometry ($B_\text{F}$), or perpendicular to it in the Voigt geometry ($B_\text{V}$). The optical signals were dispersed with an 0.5-meter spectrometer and detected by a liquid-nitrogen-cooled charge coupled device (CCD detector).

Reflectivity spectra were measured using a halogen lamp in back-reflected geometry. The exciton-polariton resonance was modeled by the approach of Ref.~\onlinecite{Hopfield1963} (details will be published elsewhere), from which the following parameters were evaluated: transverse exciton energy $E_\text{T} = 2.3220$~eV, longitudinal exciton energy $E_\text{L}= 2.3274$~eV, longitudinal-transverse splitting $\hbar\omega_\text{LT} = 5.4$~meV, and exciton damping $\hbar \Gamma = 6.7$~meV. The exciton $g$-factor of $g_\text{X}= 2.35$ was measured from the Zeeman splitting of oppositely circularly-polarized reflectivity spectra in magnetic fields up to 10~T.

The photoluminescence (PL) was excited by a continuous-wave laser with a photon energy of 2.376~eV. Low excitation densities not exceeding 10~W/cm$^{2}$ were used.

\textbf{Time-resolved photoluminescence.} The exciton recombination dynamics was measured from time-resolved PL excited with 1~ps laser pulses at 3.263~eV photon energy and detected with a streak-camera attached to an 0.5-meter spectrometer. The overall time resolution was 20~ps.

\textbf{Pump-probe time-resolved Kerr rotation.}
A polarization-sensitive pump-probe Kerr rotation technique~\cite{Yakovlev_Ch6} was employed to study the spin dynamics of carriers, for which magnetic fields up to $B_\text{V}=4$~T were applied in the Voigt geometry, i.e. perpendicular to the sample normal and to the light propagation direction. The used laser system was composed of a pulsed Ti:Sapphire laser which pumps an optical parametric oscillator (OPO) with intracavity second harmonics generation providing  wavelength-tunable emission in the range of $500-800$~nm with a spectral width of about $1$~nm and a pulse duration of 1~ps. The pulse repetition rate was 76~MHz (repetition period $T_\text{R}=13.1$~ns).

The output of the laser system was split into the pump and probe beams. The circularly polarized pump pulses create spin polarization of the carriers in the sample. The spin polarization was then analyzed by measuring the Kerr rotation of the linearly polarized probe pulses reflected from the sample. Varying the time delay between the pump and probe pulses by means of a mechanical delay line gave access to the time dependence of the spin polarization. The polarization of the pump beam was modulated between $\sigma^+$ and $\sigma^-$ by a photo-elastic modulator operated at a frequency of 84~kHz for synchronous detection.  In finite magnetic field, the Kerr rotation amplitude oscillates in time reflecting the Larmor spin precession of the carriers and decays at longer time delays. When both electrons and holes contribute to the Kerr rotation signal, as is the case for the studied CsPbBr$_3$ sample, the signal can be described with a superposition of two decaying oscillatory functions:
$A_{KR} = S_e \cos (\omega_e t) \exp(-t/T^*_{2,e}) + S_h \cos (\omega_h t) \exp(-t/T^*_{2,h})$.

\textbf{Polarization recovery measurements.} Here the pump-probe Kerr rotation was measured as well (see above), but the magnetic field $B_\text{F}$ was applied in the Faraday geometry, i.e. parallel to the sample normal and light propagation direction. In order to detect spin dynamics of the resident carriers only and to avoid the contribution of excitons, the signal was detected at a time delay of 13~ns, i.e. shortly before the arrival of the next pump laser pulse~\cite{Heisterkamp2015}. The photogenerated carrier spin polarization is stabilized by the longitudinal magnetic field, which results in an increase of the Kerr rotation amplitude with growing magnetic field. Typical polarization recovery curves (PRC) saturate in strong fields (Fig.~\ref{fig:PRC}c). The difference between the saturated Kerr rotation signal and its value at zero field is the PRC amplitude. The width of the PRC provides information on the spin relaxation mechanisms and/or on the local magnetic fields, e.g. resulting from the nuclear spin fluctuations.

\textbf{Spin inertia method.} The longitudinal spin relaxation time of the carriers, $T_1$, was measured by the spin inertia method, which is based on the pump-probe Kerr rotation technique~\cite{Heisterkamp2015}. For that, the intensity of the circularly polarized pump was modulated with frequency $f$. When the modulation period $1/f$ exceeds $T_1$, the carrier spin polarization follows the change of pump polarization. As the modulation frequency is increased, so that $1/f$ becomes comparable to $T_1$, the spin polarization cannot follow the pump polarization and the Kerr rotation modulation amplitude decreases. Additionally, a phase retardation, $\phi$, appears between the oscillations of the pump and carrier spin polarizations. One can show that for  pumping with modulated polarization in the form $R(t) = R_0 [1+\cos(2\pi f t)]$, the carrier polarization takes the form
\begin{equation}
\label{eq:SISol}
S(t)=R_0 T_1 + S_\text{ac}\cos(2\pi f t - \phi),
\end{equation}
where
\begin{equation}
\label{eq:amp}
S_\text{ac} = \frac{R_0 T_1}{\sqrt{1+(2\pi T_1 f)^2}},
\end{equation}
and
\begin{equation}
\label{eq:phase}
\quad \tan\phi = 2 \pi T_1 f .
\end{equation}
Eq.~\eqref{eq:amp} was derived in Refs.~\onlinecite{Heisterkamp2015,Smirnov2018}, while Eq.~\eqref{eq:phase} was worked out in Ref.~\onlinecite{Mikhailov2018}. One can see, that $T_1$ can be evaluated independently from the experimental dependences of $S_\text{ac}(f)$ and of $\phi(f)$. In case of a non-exponential spin dynamics the $T_1$ values determined from Eqs.~\eqref{eq:amp} and \eqref{eq:phase} will differ from each other, as the amplitude is more sensitive to the slower component, while the phase is dominated by the faster one. Note that by using the synchronous detection technique, in our experiments we detect only $S_\text{ac}$, while the time-independent component of the spin polarization [first term in the right hand side of Eq.~\eqref{eq:SISol}] is eliminated.




\clearpage

\begin{center}
{\huge Supplementary information}
\end{center}

\setcounter{equation}{0}
\setcounter{figure}{0}
\setcounter{table}{0}
\setcounter{section}{0}
\setcounter{page}{1}
\renewcommand{\theequation}{S\arabic{equation}}
\renewcommand{\thefigure}{S\arabic{figure}}
\renewcommand{\thetable}{S\arabic{table}}
\renewcommand{\thesection}{S\arabic{section}}
\renewcommand{\cite}[1]{[S\onlinecite{#1}]}

\makeatletter
    \renewcommand\@biblabel[1]{[S#1]}
\makeatother


\section{Band structure and orbital composition of Bloch functions}
\label{app:Bloch}

Bulk halide perovskite semiconductors are described by the chemical formula ABX$_3$. Here A is the organic or inorganic cation (center of the unit cell), B is the metal atom (e.g., Pb) and X is the halogen atom (e.g., Br). To be specific, we consider CsPbBr$_3$ and, for simplicity, its cubic modification described by the point group $O_h$. Note that the crystal becomes orthorhombic at low temperatures, but here we ignore the minor modifications of the selection rules and of the hyperfine interaction for different phases of the crystal. The direct band gap is formed at the R point of the Brillouin zone (the corner of the cube), the symmetry of the R valley is the same as that of the $\Gamma$ point, i.e., $O_h$.

The orbital Bloch functions of the \emph{valence band top} are invariant ($R_1^+$ representation) and of the \emph{conduction band bottom} are three functions transforming according to the $R_4^-$ representation (as vector components)~\cite{Boyer2016},\cite{Even2015}. With account for the spin and spin-orbit coupling ($R_6^+$ is the spinor representation for spin-$1/2$ states) one obtains
\begin{equation}
\label{vb}
\mbox{valence band}: \quad R_1^+ \times R_6^+ = R_6^+,
\end{equation}
\begin{equation}
\label{cb}
\mbox{conduction band}: \quad R_4^- \times R_6^+ = R_6^- + R_8^-.
\end{equation}
The bottom of the conduction band has $R_6^-$ symmetry, see Refs.~\cite{Boyer2016},\cite{Becker2018}.

According to Ref.~\cite{Boyer2016}, the valence band is mainly composed by the $s$-orbitals of the metal, $|\mathcal S_0\rangle$, with admixture of the halogen $p$-orbitals (a combination $\propto |\mathcal X_1\rangle+|\mathcal Y_2\rangle+|\mathcal Z_3\rangle$ with appropriate phase choice). For the conduction band the main contribution comes from the $p$-orbitals of the metal, $|\mathcal X_0\rangle$, $|\mathcal Y_0\rangle$, $|\mathcal Z_0\rangle$ with a slight ($\lesssim 1\%$) admixture of the $s$-orbitals of the halogen $|\mathcal S_1\rangle$.

\section{Spin polarization of resident charge carriers by circularly polarized light}
\label{app:mechanism}

The resonant absorption of circularly polarized light induces transitions from the $R_6^+$ valence band to the $R_6^-$ conduction band. Both bands are two-fold degenerate in the spin and electron-hole pairs with spin orientations $(+1/2_e,+1/2_h)$ or $(-1/2_e,-1/2_h)$ are generated, respectively, by the $\sigma^+$ and $\sigma^-$ polarized photons. The subscripts $e$ and $h$ denote electrons and holes and we use the hole representation here (the spin of the unoccupied valence band state is opposite to the hole spin).

The photocreated electron-hole pairs or excitons can transfer their spin polarization to the resident charge carriers. The polarization mechanisms were studied in detail for III-V and II-VI semiconductors, see Refs.~\cite{Yakovlev_Ch6},\cite{Glazov2012} for review. The mechanisms are related to either the formation of bound three particle complexes (positively or negatively charged excitons, also termed as trions) or to the exchange scattering of the excitons by the resident carriers. Particularly, in the latter case the scattering of the spin-polarized exciton with an unpolarized electron and hole results in the exchange of the identical carriers and in the transfer of the spin polarization to the resident electrons or holes. After exciton recombination the spin polarization remains in the system of resident carriers.

\section{Hyperfine interaction in perovskites}
\label{app:HF}

For localized carriers the spin-orbit coupling effects are suppressed due to the quenching of the orbital motion. Thus, the hyperfine coupling of electron or hole spins with host lattice nuclei are the prime candidates for driving the spin dynamics~\cite{OO1984}. The nuclear spins experience the Knight field from the spin polarized charge carriers and, in turn, the charge carriers experience the Overhauser field induced by the nuclei.

The hyperfine interaction between the charge carriers and nuclei provides a transfer of spin angular momentum from the electrons or holes to the nuclei and thereby results in a dynamical nuclear polarization. Generally, the hyperfine coupling Hamiltonian can be written in the phenomenological form of a scalar product of the nuclear and charge carrier spins (see below for details):
\begin{equation}
\label{hf:gen}
\mathcal H_{hf} = A_{e(h)}v_0 |\varphi_{e(h)}(\bm R)|^2  (\mathbf I \cdot \mathbf S_{e(h)}),
\end{equation}
where $\mathbf I$ is the nuclear spin, $\mathbf S_e$ ($\mathbf S_h$) is the electron (hole) spin, $A_e$ ($A_h$) is the corresponding interaction constant. The factors $v_0$, the unit cell volume, and $|\varphi_{e(h)}(\bm R)|^2$, the absolute square of the envelope function of the charge carrier at the nucleus position, are introduced in Eq.~\eqref{hf:gen} to make $A_{e(h)}$ dimensionless. The anisotropic terms allowed in cubic semiconductors for $I>1/2$ are usually small and disregarded hereafter.

In the presence of an external magnetic field, $\mathbf B$, the hyperfine interaction [Eq.~\eqref{hf:gen}] provides the spin transfer between electron (hole) and nuclear spins. In the experimental configuration shown in Fig.~3(b) in the main text there is a non-zero component of the charge carrier spin $\mathbf S_{e(h)}$ onto the magnetic field $\mathbf B$. Flip-flop hyperfine processes give rise to the dynamical nuclear spin polarization $\langle \mathbf I \rangle$ in the form~\cite{OO1984}
\begin{equation}
\label{dnp}
\langle \mathbf I \rangle = \ell_{e(h)} \frac{4I(I+1)}{3} \frac{\mathbf B(\mathbf B\cdot \mathbf S_{e(h)})}{B^2},
\end{equation}
where $\ell_{e(h)}\leq 1$ is the leakage factor characterizing the losses of nuclear spin polarization due to relaxation processes other than the hyperfine coupling.

Via the hyperfine interaction, the polarized nuclear spins produce the Overhauser field
\begin{equation}
\label{Over:gen}
\mathbf B_{\mathrm N,e(h)} = (g_{e(h)} \mu_B)^{-1}\sum_j A_{e(h)}^j v_0 |\varphi_{e(h)}(\bm R_j)|^2  \langle \mathbf I_j \rangle,
\end{equation}
where the summation is carried out over all nuclei, so that the index $j$ includes all chemical elements, all isotopes of the element abundant in the sample, as well as all positions $\bm R_j$ of the nuclei. Under the standard assumption of a uniform nuclear spin polarization $\langle \mathbf I \rangle$ Eq.~\eqref{Over:gen} can be written in a simple form as
\begin{equation}
\label{Over:gen:1}
\mathbf B_{\mathrm N,e(h)} = (g_{e(h)} \mu_B)^{-1}\sum_i A_{e(h)}^i \langle\mathbf I_i \rangle,
\end{equation}
where the sum is carried out over the different elements and isotopes denoted by the subscript $i$. Indeed, the summation over unit cells assuming homogeneous  nuclear polarization, can be transformed to an integral as
\begin{equation}
\sum_{j} v_0 |\varphi_{e(h)} (\bm R_j)|^2 = N_{iso} \int d\bm R_j  |\varphi_{e(h)} (\bm R_j)|^2 = 1,
\end{equation}
where $N_{iso}$ is the number of corresponding isotopes in the unit cell.  In line with the smooth envelope method, we assume that $\varphi_{e(h)} (\bm R_j)$ does not vary much on the scale of the lattice constant.

The Overhauser field $\mathbf B_{\mathrm N}$ adds up to the external field changing the frequency of the carrier spin precession. The direction of $\mathbf B_{\rm N,e(h)}$ is determined by the sign of the hyperfine coupling constant $A_{e(h)}$ and the direction of $\mathbf I$ which, in turn, is governed by $\mathbf S_{e(h)}$, i.e., it can be adjusted by varying the light helicity. We performed measurements for both $\sigma^+$ and $\sigma^-$ circularly polarized pumping, inducing the opposite directions of $\mathbf B_{\rm N,e(h)}$. Measuring the change of the electron and hole spin precession frequencies we detected the nuclear field.

Figure~3b in the main text shows that the spin precessions for opposite pump polarizations acquire a small, but still reliably detectable phase shift which increases with time delay (inset to Fig.3b in the main text). This means that the precession frequencies for $\sigma^+$ and $\sigma^-$ pumping are different. An accurate fit to the experimental data shows that the  nuclear field acting on the electron spins is $|\mathbf B_{\mathrm N,e}| = 1.0 \pm 0.8$~mT and that on the hole spins is $|\mathbf B_{\mathrm N,h}| =3.1 \pm 0.5$~mT. We conclude that the hyperfine interaction mostly affects the valence band hole spins.

This experimental result may seem surprising compared with the widely studied III-V and II-VI semiconductors where the hyperfine coupling is dominated by the conduction band electrons~\cite{Paget1977},\cite{Gryncharova1977},\cite{Chekhovich2013},\cite{Chekhovich2017}. Our theoretical analysis, however, confirms that in  perovskite like CsPbBr$_3$ the hyperfine coupling for the valence band holes is stronger compared to the conduction band electrons. To that end we consider in more detail the atomic orbital composition of the Bloch states for electrons and holes. The valence band states in the vicinity of the $R$-point of the Brillouin zone, where the direct band gap is formed, is mainly composed by $s$-type atomic orbitals of the metal (lead in our case) with an admixture of $p$-type atomic orbitals of the halogen (i.e., bromine), see, e.g., Ref.~\cite{Boyer2016} and Appendix~\ref{app:Bloch} for details. The leading contribution to the hyperfine coupling is provided by the Fermi contact interaction for the $s$-type orbitals, which does not vanish at the positions of the nuclei. By contrast, the conduction band is mainly formed from the $p$-type metal orbitals with a slight admixture of the $s$-type orbitals of the halogen. The magnetic dipole-dipole interaction for $p$-type orbitals is about an order of magnitude weaker~\cite{Gryncharova1977},\cite{Chekhovich2017}.

The hyperfine interaction constants for CsPbBr$_3$ can be estimated as follows. In the \textbf{valence band} the hyperfine coupling is dominated by the contact interaction with the the lead atoms. The constant $A_h^{(0)}$ calculated per isotope, i.e., disregarding the abundance, can be written as~\cite{Chekhovich2013}:
\begin{equation}
\label{A}
A_{h}^{(0)} = \frac{16\pi \mu_B \mu_I}{3 I} v_0^{-1} |\mathcal S_0(0)|^2,
\end{equation}
where $\mathcal S_0(\bm r)$ is the Bloch function at the nucleus position normalized per volume of the unit cell, $\int_{v_0} |\mathcal S_0(\bm r)|^2d\bm r = v_0$, $\mu_B$ is the Bohr magneton, $\mu_I$ is the nuclear magnetic moment, $I$ is the spin of the nucleus. It is important to note that the transformation from the electron to the hole representation results in the inversion of both the direction of spin and the energy axis, leaving the hyperfine constant sign the same. That is why we can use the electron representation for evaluation of the hyperfine coupling for holes.

For holes the relevant isotope is $^{207}$Pb with an abundance of about $22\%$, the nuclear spin $I=1/2$ and $\mu_I=0.58\mu_N$, where $\mu_N\approx 7.62$~MHz/T is the nuclear magneton. The hyperfine interaction can be estimated from the atomic constants~\cite{Morton1978},\cite{Koh1985}. From Ref.~\cite{Morton1978} (which uses an approach that typically overestimates the value, as known from the comparison for III-V semiconductors) we have (per nucleus):
\begin{equation}
\label{Mort:Pb}
A_{h}^{(0)} = 107~\mu\mbox{eV}.
\end{equation}
Note that inclusion of the so-called Mackey-Wood correction gives a $\sim 3.15$-fold enhancement up to $336~\mu\mbox{eV}$.
From Ref.~\cite{Koh1985} (which approach typically underestimates the hyperfine coupling), disregarding the anisotropy factor we have :
\begin{equation}
\label{Koh:Pb}
A_{h}^{(0)} = 78~\mu\mbox{eV}.
\end{equation}
Thus, we take $A_{h}^{(0)} \approx 100$~$\mu$eV as a conservative estimate which is also in agreement with estimates and measurements of the Fermi contact interaction in Pb$_{1-x}$Sn$_x$Te~\cite{Hewes1973}. Taking into account that the natural abundance of $^{207}$Pb is $\beta\approx 22\%$~\cite{Fuller1976}, we finally have $A_{h}\approx 20$~$\mu$eV.

Note that the dipole-dipole interaction with $^{79}$Br and $^{81}$Br (both $I=3/2$) can be estimated as $A_{h}^1 \sim |C_p|^2\times 7$~$\mu$eV, where $C_p$ ($|C_p|\ll 1$) is the admixture of the $p$-shell of Br to the $s$-shell of Pb in the valence band Bloch function, therefore, it can be disregarded.

For the \textbf{conduction band} states it is sufficient to account for the dipole-dipole interaction with the bromine nuclei only. Estimates based on Refs.~\cite{Morton1978},\cite{Koh1985} give $A_{e} \approx 7$~$\mu$eV. We use this constant for both the $^{79}_{35}$Br and $^{81}_{35}$Br isotopes whose spin is the identical and whose total abundance is $100\%$. The parameters of the hyperfine interaction are summarized in Tab.~S1.

\begin{table}[t]
\caption{Parameters of the hyperfine interaction in CsPbBr$_3$. The values of $A_e$ and $A_h$ are given following the estimates in the text, the values in parenthesis are normalized to the abundance. Only isotopes with significant abundance are included.
}
\label{tab:isotopes}
\begin{tabular}{c|c|c|c|c|c}
\hline
Isotope & Spin & $\mu_I/\mu_N$ & Abundance & $A_e$  & $A_h$ \\
 &  & & $\beta$ & ($\mu$eV) & ($\mu$eV) \\
\hline
$^{133}_{55}$Cs & $7/2$ & $2.58$  & $100\%$ & $-$\footnote{The contribution of its orbitals to the Bloch function in negligible.} & $-^{a}$ \\
$^{206}_{82}$Pb & $0$ & $0$ & $24.1\%$ & $-$\footnote{This isotope has spin $0$ and the hyperfine interaction is absent. } & $-^{b}$ \\
$^{207}_{82}$Pb & $1/2$ & $0.58$ & $22.1\%$ & $-^a$ & $100$ ($20$) \\
$^{208}_{82}$Pb & $0$ & $0$ & $52.4\%$ & $-^b$ & $-^{b}$ \\
$^{79}_{35}$Br & $3/2$ & $2.1$ &$50.7\%$ & $7$ ($3.5$) & $-^{a}$ \\
$^{81}_{35}$Br & $3/2$ & $2.27$ &$49.3\%$ & $7$ ($3.5$) & $-^{a}$ \\
\hline
\end{tabular}
\end{table}

For complete nuclear polarization, where $|\langle \mathbf I\rangle|=I$ according to our estimates after Eq.~\eqref{Over:gen:1} the maximum Overhauser fields read
\begin{equation}
\label{maximal}
|B_{\mathrm N,h}^{max}| \approx 230~\mbox{mT}, \quad |B_{\mathrm N,e}^{max}| \approx  280~\mbox{mT}
\end{equation}
for the holes and electrons, respectively. Interestingly, the maximal Overhauser field is slightly larger for the conduction band despite the smaller hyperfine coupling constants. This is because there are three Br nuclei in the unit cell with $I=3/2$ and because of the smaller abundance of Pb. However, the dynamical nuclear polarization process is controlled by individual spin-flips, where the hyperfine interaction with holes prevails. The flip-flop of Br nuclei with the electron spin is controlled by the much smaller individual interaction constant. Thus, we expect the dynamical nuclear polarization of Br to be weaker, in agreement with the smaller leakage factors obtained in the experiment: $\ell_h \sim 0.08$ for the valence band and $\ell_e \sim 0.02$ for the conduction band. We took the leakage factors from the measured Overhauser field using the estimates in Eqs.~\eqref{dnp} and \eqref{maximal} and considering the angle of $\sim 75^{\circ}$ between the magnetic field and the light propagation axis.

\section{Zeeman Hamiltonian and $g$-factor of carriers and excitons}
\label{app:Zeeman}

We define the effective exciton Zeeman Hamiltonian in the form \cite{Kiselev1996},\cite{Durnev2012},\cite{Yu2016},\cite{Ramade2018}
\begin{equation}\label{Zeeman}
\widehat{H}_{Z}=\mu_Bg_e\left(\bm {S}_e \bm B\right) + \mu_B g_h\left( \tilde{\bm S}_h\bm B\right).
\end{equation}
Here $\mu_B$ is the Bohr magneton, $g_e$ and $g_h$ are the electron and hole $g$-factors, respectively, $S_e=\tilde{S}_h=\pm{1/2}$ are the electron and hole spins. Note that we use the standard representation for the exciton Hamiltonian where the electron part is taken in the electron and the hole part is taken in the hole presentation, that is, why the Hamiltonian is the sum of the electron and the hole contributions.
In this definition the $g$-factors are positive both for electrons and holes if the ground state is the state with spin projection $s_z=-1/2$ onto the direction of the magnetic field. Note that just like for the hyperfine coupling, the value and the sign of the hole $g$-factor is the same, both for the electron and hole representations. The exciton $g$-factor is defined as
\begin{equation}\label{gfactor}
g_X=\frac{E_{\sigma^+ }-E_{\sigma^-}}{\mu_B B}.
\end{equation}
In case of the perovskites the exciton $g$-factor is the sum of the electron and hole $g$-factors: $g_X=g_e+g_h$. The slight discrepancy between the values of the $g_X$ inferred directly from optical spectroscopy, on the one hand, and from the sum of $g_e$ and $g_h$, on the other hand, may be attributed to the effects of the electron-hole exchange interaction and band non-parabolicity which gives rise to a renormalization of the $g$-factors of the charge carriers in the exciton~\cite{Grigoryev2016}.

\section{Analysis of the spin inertia data and polarization recovery curves}
\label{app:PRC}

\begin{figure}
    \centering
\includegraphics[width=\columnwidth]{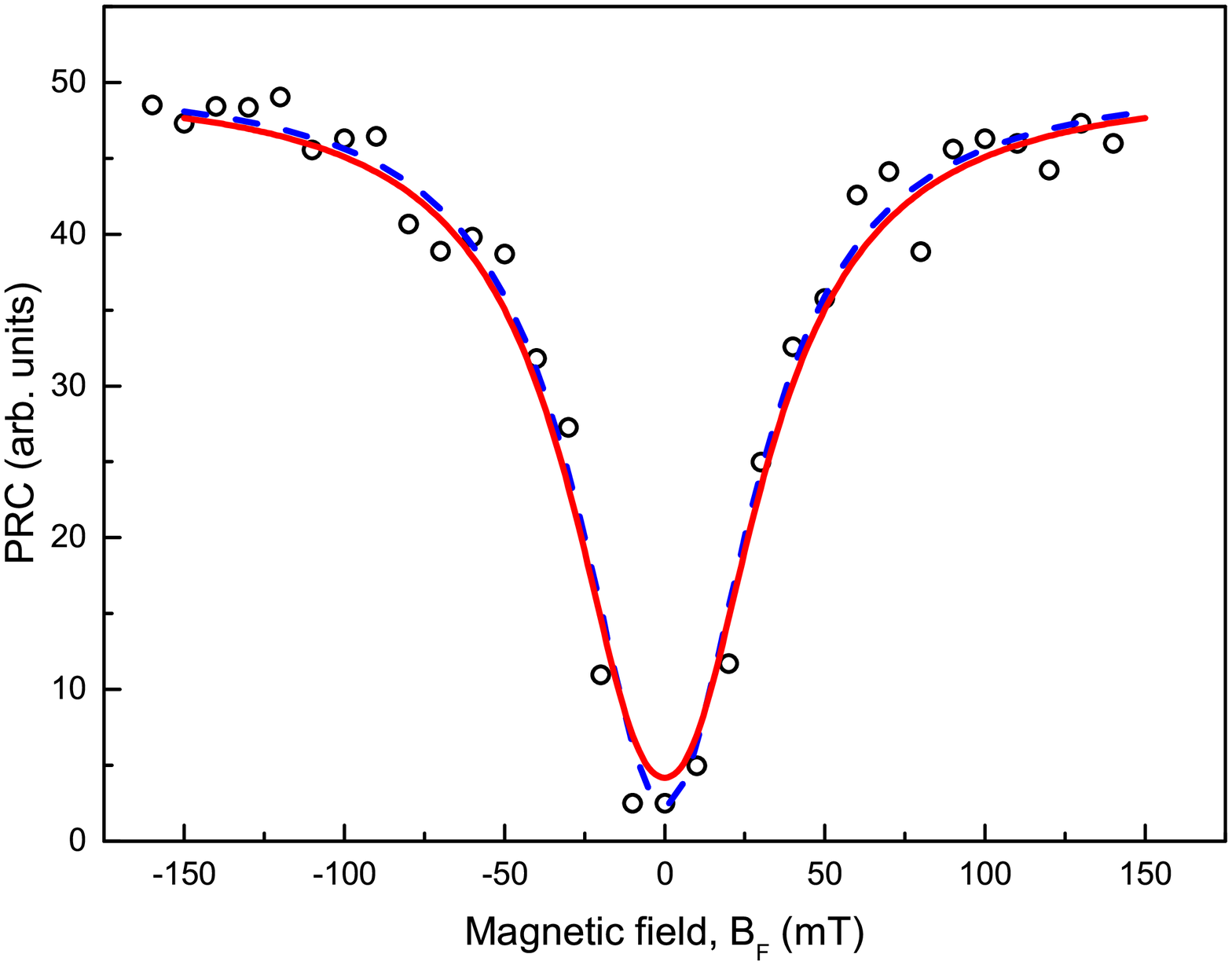}
\caption{Polarization recovery curve (PRC) measured at lowest modulation rate of 1~MHz together with fit after simplified Eq.~\eqref{T1fit} (blue dashed line) and results of calculation after the full model of Ref.~\cite{Smirnov2018} (red line), see text for parameters.
}
    \label{fig:PRCfit}
\end{figure}

In order to analyze the magnetic field dependence of the polarization recovery curves we employ the model developed in Ref.~\cite{Smirnov2018}. It accounts for both the spin precession in the field of the nuclear fluctuations and the finite correlation time of the hole at a localization site, $\tau_c$. We consider the case of small modulation frequencies where the spin inertia is not important. Introducing $\omega_N$ as the characteristic fluctuation of the Overhauser field we obtain
\begin{equation}
\label{T1fit}
PRC \propto \left(\frac{1}{\tau_s} + \frac{\omega_N^2\tau_c}{1+(\omega_{\mathrm L,h}\tau_c)^2} \right)^{-1}.
\end{equation}
We recall that $\omega_{\mathrm L,h}$ is the hole spin precession frequency in the external magnetic field and $\tau_s$ is the spin relaxation time unrelated to the hyperfine coupling. Equation~\eqref{T1fit} holds provided $\omega_N \tau_c \ll 1$ or $\omega_{L,h} \gg
\omega_N$. The fit after Eq.~\eqref{T1fit} with $\tau_c$ and $\omega_N$ being free parameters and $\tau_s=50$~ns (inferred from the spin inertia measurements) gives the correlation time $\tau_c\approx 2.1$~ns and the nuclear field fluctuation $\delta B_N = \hbar \omega_N/(g_h\mu_B) \approx 6.6$~mT. In this case $\omega_N \tau_c \approx 0.9$ and Eq.~\eqref{T1fit} is not applicable at $B \lesssim \delta B_N$. Thus, to confirm the results, we have performed the full calculation after the model of Ref.~\cite{Smirnov2018} shown by the red curve in Fig.~\ref{fig:PRCfit} for the same parameter values. Very good agreement between the calculations and experimental data is observed.

The value of the nuclear spin fluctuation $\delta B_N \approx 6.6$~mT allows us to make a crude estimate of the hole localization length. To that end we evaluate $\omega_N$, assuming independent contributions of the Pb isotopes as
\begin{equation}
\label{omega_N_est}
\omega_N =\sqrt{\frac{2}{3\hbar^2}I(I+1)}\frac{A_h^{(0)}\sqrt{N_{\rm Pb}}}{N_{cells}},
\end{equation}
where $N_{cells}$ is the number of cells within the hole localization volume, $N_{\rm Pb} = \beta N_{cells}$ is the number of lead isotopes with non-zero spin within the hole localization volume. It follows from Eq.~\eqref{omega_N_est} that for the experimental value of $\omega_N = g_h\mu_B  \delta B_N/\hbar \approx 0.44$~ns$^{-1}$ the localization volume contains $N_{cells} \approx 8\times 10^3$ unit cells, which gives the localization length $L_{loc} = a_0 N_{cells}^{1/3} \approx 11.5$~nm.

\newpage

\end{document}